# The Baltic Meetings 1957 to 1967

Erik Høg   - Niels Bohr Institute, Copenhagen - ehoeg@hotmail.dk

**Abstract:** The Baltic meetings of astronomers from Northern Germany and Scandinavia began in 1957 and gathered up to 70 participants. Reports of the presentations are available from all meetings, providing an overview of the interests of astronomers in this part of the world 50 years ago. Most interesting to see for a young astronomer in our days, I think, is that a large part of the time was about astrometry. This focus on astrometry was the basis for the scientific knowhow which made the idea of space astrometry realistic, resulting in the approval by ESA of the first astrometry satellite Hipparcos in 1980 which brought a revolution of high-precision astrometry of positions, motions and distances of stars. The correspondence with ten observatories shows that only one of them has any archive of letters at all from the 1950s, that is in Copenhagen where about 7000 letters on scientific and administrative matters are extant.

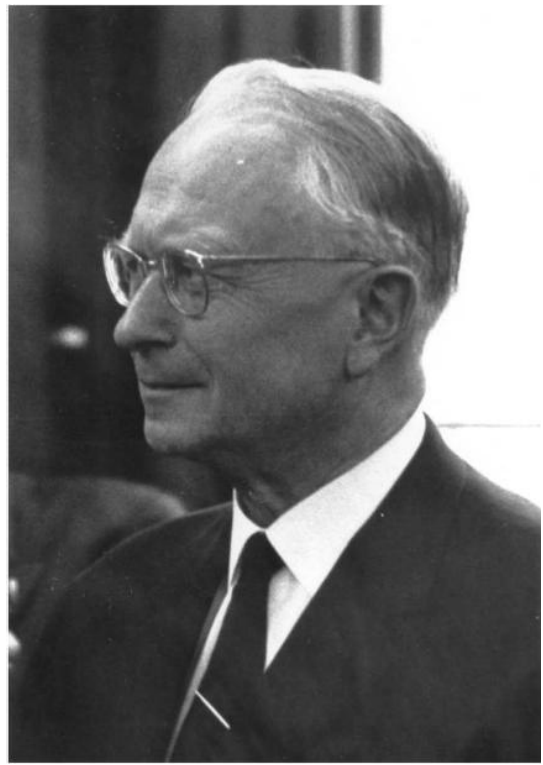

Figure 1   Otto Heckmann. – Photo: Wilhelm Dieckvoss

# 1 Introduction

There were meetings in the 1950-60s, then sometimes called Baltic Meetings, in Aarhus, Copenhagen, Lund, Hamburg, and Göttingen, five in total between 1957 and 1967. They resulted from the initiative of professor Otto Heckmann (1901-83) (director of the Hamburg-Bergedorf observatory 1942-62, and later president of the Astronomische Gesellschaft, president of the International Astronomical Union (IAU) and the first director general of ESO). Heckmann wanted to build a contact to Scandinavian astronomers after the war. Participants came from Hamburg, Göttingen, Bonn, Kiel, Copenhagen, Aarhus, Oslo, Lund, Stockholm, and Uppsala at these meetings: Dieckvoss, Larink, and Wellmann from Hamburg, Behr and Schmidt-Kaler from Göttingen, Meurers from Bonn, perhaps Unsöld from Kiel, Baschek from Kiel, Rosseland from Oslo, Schalén, Larsson-Leander and Hansson from Lund, Bertil Lindblad from Stockholm, Holmberg and Malmquist from Uppsala were there. Up to 70 participants attended sometimes.

Otto Heckmann visited the very new Brorfelde Observatory on 18. September 1957. He stayed with the Peter Naur (*1928) family over night and visited Copenhagen Observatory on the 19. before going to the meeting in Lund on 19.-21. September.

My first memory of this is a small meeting in Copenhagen on 19. September 1957. It was Otto Heckmann's first visit to Copenhagen to establish a contact. There were just a few persons in Julie Vinter Hansen's (1890-1960) apartment in the observatory at Østervold that day. I remember Peter Naur being there, he was my mentor and encouraged me to ask Heckmann a question related to my work at the new meridian circle in Brorfelde, a question which we had been discussing.

One of my young colleagues in Bergedorf told me later that Heckmann had reported of his first visit to Copenhagen. He had met a young guy working with the meridian circle in Brorfelde and *"he seemed not to know really what he was doing"*, my colleague told. I had shown Heckmann a series of measures from photographic observations of the Polarissima, a star very close to the north celestial pole. I had taken them with the new meridian circle in Brorfelde in order to monitor the stability of the instrument. There were only some fifteen measures in one coordinate and I was pondering whether one could say that there was a drift or not. I believe Heckmann advised me to get some more measures. That was the first impression Heckmann had of me, the *"young guy"* on whom he placed a very great responsibility three years later.

# 2 Meeting in Lund 1957

The first Baltic meeting was gathered in Lund on the 19.-21. September 1957. A report appeared in Swedish in PAT (1957). In the following extract I will mention the speakers and their subjects, the intention being to render an impression of the astronomical interests 50 years ago. In the extract, I will generally omit Dr., Professor, laborator, docent, assistent and other titles which appear throughout the Swedish text. Neither will I mention all dinners, speeches and receptions.

The director of Lund Observatory, professor Carl Schalén (1902-93), opened the meeting of about 40 German, Danish and Swedish astronomers, there were 10 participants from Hamburg. He explained that the idea at first was to gather astronomers from the nearby observatories in Copenhagen, Aarhus, Hamburg-Bergedorf, Kiel and Lund for presentations and discussions. Later on the circle had been expanded to Saltsjöbaden and Uppsala in the north and to Göttingen and Bonn in the south. Nobody from Kiel was able to attend this time. In the Astronomische Jahresberichte for 1957 is mentioned "a working meeting in September in Lund where Swedish, Danish and German astronomers participated, with ten from Hamburg."

The presentations began in the morning of 20. September and were mainly about interstellar matter and positional astronomy. Knut Lundmark (1889-1958), Lund, spoke about his catalogue of nebulae and studies of novae in connection with absorption. A. Behr and Th. Schmidt-Kaler from Göttingen presented their observations of the polarization of light at its passage through the interstellar medium. Behr's results were related to nearby stars and were obtained with his very accurate instrument built in the observatory. Schmidt-Kaler had observed about 30 Cepheids.

In the afternoon the electronic computer Smil was demonstrated by C.-E. Fröberg. Ingrid Torgård, also from Lund, spoke on the reduction of meridian observations using Smil. A visit to "Fysiska institutionen" was arranged by Bengt Edlén. The participants showed great interest both in Smil and in the Fysikum; the computer is often called SMIL, with capitals, but not in PAT (1957), surprisingly.

At 6 pm the meeting started again with J. Meurers, Bonn, speaking about motions in the open cluster NGC 1960 and the problem of stellar associations. E. Brosterhuis, Bergedorf, spoke about his use of the Schmidt telescope for astrometry. The newly renovated meridian circle was demonstrated by A. Reiz and N. Hansson.

The next day started with a lecture by Erik Holmberg, Uppsala, about his new photometric studies of galaxies. P. Wellmann, Bergedorf, spoke about spectral classification and described his detailed studies of F stars with high dispersion. G. Larsson-Leander spoke about absorption in the direction of the cluster NGC 1664 and also about the anomalous tail of comet Arend-Roland.

The afternoon began with a lecture by K. Eberlein, Bergedorf, about infrared photometry in the Praesepe cluster. J. Ramberg spoke on his studies of the distribution of stars in the Milky Way. J. Hardorp, Bergedorf, spoke on colour-magnitude diagrams of clusters and critized the fine-structure claimed by Eggen and Pilowski. I. Oszvath, also Bergedorf, spoke on RR Lyrae stars in M 3.

Most of the presentations were followed by a lively discussion.

The "newly renovated meridian circle" presented above by Reiz and Hansson never produced very important observations but it played a crucial role in the history of astrometry many years later. It stood there with it fine mechanics and fascinated the young Lennart Lindegren (1950-) as he explained me at my visit in 1973 (Høg 2008). Without the old renovated meridian circle Lennart Lindegren would probably have been more attracted by astrophysics and without Lennart there would have been no approval of Hipparcos in 1980, according to Høg (2011a). Without Lennart there would have come no Gaia mission either, see Høg (2011b).

The conference ended with a dinner where the prorektor of the Lund University was present and where O. Heckmann and Julie Vinter-Hansen thanked on behalf of respectively the Germans and the Danes. The intention to repeat this kind of symposia in the future was expressed.

## 3  Hamburg-Bergedorf in 1958

During his visit in Denmark, Heckmann promised to place the Brorfelde Observatory on the list to receive the Bergedorf publications. Heckmann invited Peter Naur (1928-) and Kjeld Gyldenkerne (1919-99) to give colloquia in Bergedorf which were always held on Saturdays, according to the following programme for January and February 1958 where four out of eight presentations had an astrometric subject.

"Astronomisches Colloquium auf der Hamburger Sternwarte – Sonnabends 10.45 Uhr"
11.1.1958: P. NAUR  – Tollose, Studies of the motion of minor planet 51 Nemausa.
18.1.1958: A VELGHE – Uccle, The search for relatively cool stars, with Schmidt telescopes.
25.1.1958: E. SCHÜCKING, Entstehung der Elemente nach L. Heller und anderen.
1.2.1958: J. LARINK, Der Nullpunkt des Fundamentalsystems der Deklinationen nach P. Naur.
  W. DIECKVOSS, Eigenbewegungsarbeiten von S. Vasilevskis.
8.2.1958 K. GYLDENKERNE – Tollose, Quantitative methods for spectral classification applicable in galactic research.
15.2.1958 v.d. HEIDE, Über Zeitbestimmungen.
22.2.1958 K. BAHNER, - Heidelberg, Photometrie galaktischer Cepheiden.

As so often Dr. Wilhelm Dieckvoss took pictures of both Naur and Gyldenkerne, but only the one of Naur is available to me, Figure 2. Unforgettable was also that nobody ever smoked in the presence of Heckmann - in those times where smoking everywhere was normal.

Naur wrote a very polite letter of thanks to Heckmann, dated 19. Jan. 1958. Interesting in this letter is the comment on automatic instruments: *"…It was very interesting and inspiring to see your great and active observatory. I have been a little amused at the many automatic devices attached to the various instruments. I had gathered the impression that you are somewhat sceptic about too much of this kind. Perhaps you have arrived at this point of view from actual experience with the Bergedorf instruments?!"*

Heckmann's answer is not available but we are coming closer to understand the matter of automation at the Hamburg observatory through my letter to Naur of 18.12.1958 from which I will therefore quote. In the letter I am first telling about my work, about the common coffee in the Schmidt Telescope building, and about the weekly seminars on Martin Schwarzschild's book on stellar evolution. The weekly colloquia had contained two reports from the IAU assembly in Moscow, Wurm had spoken on his observation of galactic nebulae, and (translated from Danish). Then follows on automation: *"Last Saturday I reported on automatic measurement in astronomy, including my own ideas in that direction [see my report elsewhere from 2015: "Young astronomer in Denmark 1946-58"]. – Some of the young and also the elder, especially Wellmann are interested in automation. People here say that Heckmann is not interested and that he ridicules the young who speculate about automatic machines instead of measuring stars. I cannot complain, however, about his interest in me and my ideas, but he does not have to account for me to anybody. So he does not have to care whether I am wasting my time. As far as I can see, his attitude is that he does want automation, but he does not want experiments, it must come at a slow and safe speed.*

*Such an attitude can however mean that people interested in automation will disappear. I know that Dr. Tripp leaving for America in February. Dr. Tripp (about 30 years) is secretary for ESO (European Southern Observatory), about this rather depressing enterprise I will however not elaborate here. Tripp has read my report about automation, so we immediately understood each other."*

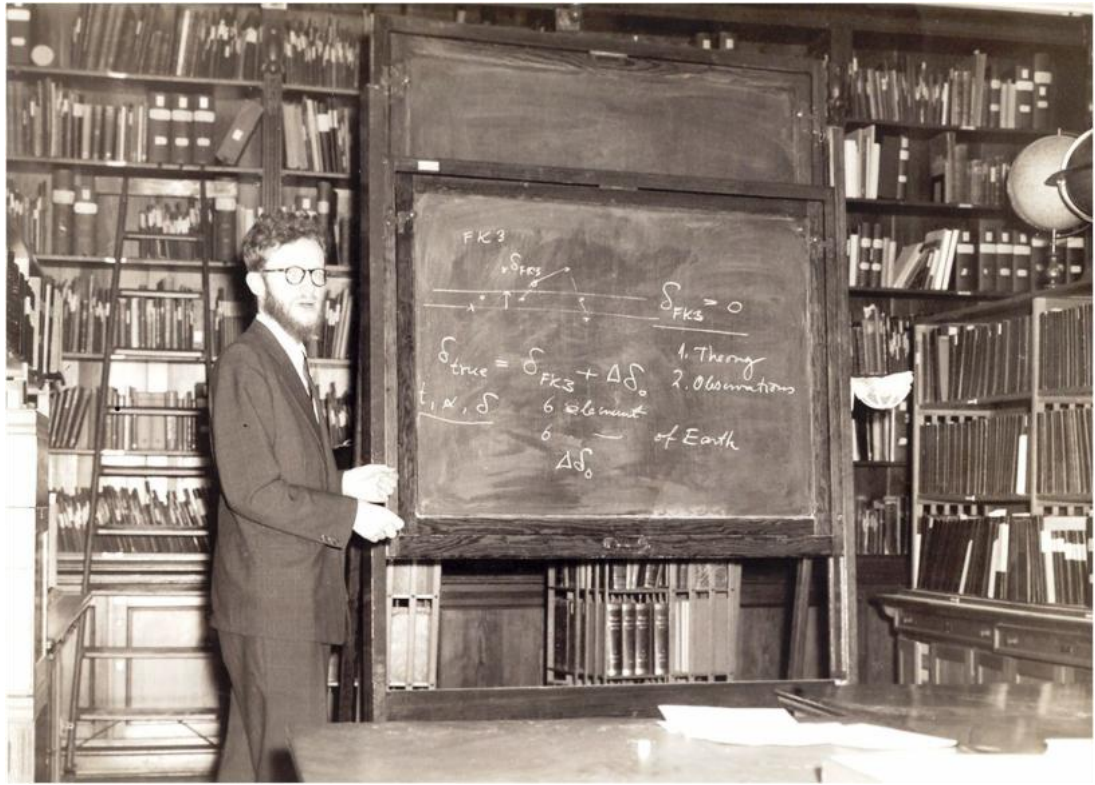

Figure 2 Peter Naur in the Bergedorf library in January 1958 giving a talk on an astrometric subject, systematic errors in the declination system. Chalk on blackboard was the standard presentation technique before the revolutions came with overheads and then with PowerPoint. – Photo: Wilhelm Dieckvoss.

I continue my letter with more on my work with spectro-photometry on Schmidt plates, measurement of double stars with the 60 cm refractor, and about the computers Zuse 22 and IBM 650.

In 1959 first steps were taken in direction of automation. I was allowed to mount digitizers on the iris photometer and the spectrum scanner, see Figure 4 in Høg (2014a), a low budget project where I could design the electronics and do the soldering. The data were punched into cards with a standard IBM machine and especially the photometer was used daily to measure the plates of galactic clusters from the Schmidt telescope. This equipment became a showpiece for foreign guests and Heckmann' colleagues from the faculty, also a small delegation from DESY (Deutsches Elektronen Synchrotron) came on his invitation. This paved the way for Heckmann's immediate acceptance of my ideas in July 1960 about photon counting astrometry using punched tape to record the observations with the meridian circle, see Section 3 of Høg (2014a).

## 4  Meeting in Aarhus 1958

There was a meeting in Aarhus 1.-2. September 1958. But it does not quite belong to the series of Baltic meetings because the following one in Göttingen is called the second by Heckmann in his report. Nevertheless, I repeat the message received from Svend Laustsen (1927-). He does not remember any report, only that Naur and Rudkjøbing were engaged in a very lively discussion about the nature of aberration. But he is not able to explain what the issue was. – Even though this meeting does not belong in the series, it could shed light on life among astronomers 50 years ago, especially if I receive other information about it.

Peter Naur remembers very well this meeting in Aarhus of Swedish and Danish astronomers and he has sent a draft agenda from the organizer, Mogens Rudkjøbing. Ten speakers are listed: Gyldenkerne, Kristenson, B.-A. Lindblad, Hansson, Lyngå or Roslund, Axel Nielsen, Reiz, Strömgren, Rudkjøbing, Naur. Naur remembers the discussion with Rudkjøbing about aberration and he has sent two pages on this matter, included here as Naur (2015). Naur's note is an account of controversies on astronomical aberration such as he has been aware of them during the years 1956 to 1995 and references are included.

## 5  Meeting in Göttingen 1959

The second meeting was held in Göttingen 16.-17. March 1959. The guest book of the Göttingen Observatory in Figure 3 shows the signature of 39 participants.

O. Heckmann distributed in January 1960 a report of 36 pages with summaries of the 20 presentations, typed and copied in the institute. His introduction: "Das vorliegende Heft enthält die Zusammenfassungen der Vorträge des zweiten gemeinsamen Colloquiums einiger skandinavischer und deutscher Sternwarten im März in Göttingen. – Die Vorträge sind widergegeben in der Reihenfolge, in der sie gehalten wurden."

Die Vorträge am 16. März 1959: <u>Die Struktur des galaktischen Systems</u>

1. E. Holmberg - Lund: On the masses of extragalactic nebulae
2. P.O. Lindblad - Stockholm: Mutual perturbations of ring formations in galaxies
3. A. Elvius - Uppsala: A possible influence of magnetic forces on the development of spiral arms
4. B.H. Grahl - Bonn: Messprofile der 21 cm-Linie
5. J. Hardorp - Hamburg: Durchmusterung der Milchstrasse nach Sternen hoher Leuchtkraft
6. K. Rohlfs - Hamburg: Spektral- und Leuchtkraftkriterien bei G- und K-Sternen
7. K.G. Malmquist - Uppsala: Absorption and space-distribution
8. C. Schalen - Lund: Über die interstellare Verfärbung im Cygnus-Gebiet
9. A. Behr - Göttingen: Interstellare Polarisation des Sternlichts in Sonnenumgebung
10. G. Lyngå - Lund: Photoelectric measurements at the Boyden Observatory
11. B. Ljunggren - Uppsala: The colours of stars near the north galactic pole
12. I. Ozsvath - Hamburg: Photometrie des offenen Sternhaufens NGC 7789

Figure 3. The guest book from Göttingen observatory at the meeting on 16.-17. March 1959. - Photo: Courtesy of Institute of Astronomy Göttingen (IAG).

Die Vorträge am 17. März 1959: <u>Astronomische Instrumente</u>

1. K.H. David - Göttingen: Das lichtelektrische Photometer am Göttinger Sonnenturm
2. I. Hiller - Hamburg: Ein Plattenphotometer zur Registrierung von Intensitäten
3. G. Brückner - Göttingen: Ein Intensitäten registrierendes Plattenphotometer
4. E. Hóg - Kopenhagen/Hamburg: A study in automatic spectrophotometry
5. W. Dieckvoss - Hamburg: Zeitliche Änderungen astrometrischer Objektive
6. B.H. Grahl - Bonn: Justierungsmethoden für Radioteleskope
7. H. Schmidt - Bonn/Hoher List: Bemerkungen zur Frage der Verwendung von Schmidt-Teleskopen für photometrische Zwecke
8. P. ten Bruggencate - Göttingen: Das Institut für Sonnenforschung in Locarno-Monti

## 6  Meeting in Nijenrode 1963

A "European Astronomers Conference" July 29 – August 3, 1963 in Nijenrode, Holland, gathered participants from USA, Holland, France, Sweden, Denmark and elsewhere. It does not belong to the Baltic meetings.

## 7  Meeting in Copenhagen 1962

The third Baltic meeting took place in Copenhagen and Brorfelde 14.-15. April 1962. A report appeared in Danish in NAT (1962). The meeting was attended by 70 astronomers from Scandinavia and Northern Germany, as listed in Copenhagen (1962). Saturday 14. April was dedicated to lectures and discussions while the Sunday was used for an excursion to the Brorfelde observatory.

Saturday before noon was dedicated to astrometry with Heckmann in the chair. Professor Svein Rosseland (1894-1985), Oslo, chaired the astrophysical presentations in the afternoon. Short reports of the presentations follow on 4 pages in Danish, all based on a longer report which "will be sent" to the participants – Whether this was in fact done I do not know. I will here extract only the name of the speaker and the title of the talk.

The presentations were:

G. van Herk: Can the efficiency of astrometric work be increased?

P. Stumpf: Durchgeführte und geplante Arbeiten aus dem Gebiet der Fundamentalastronomie mit Hilfe der elektronischen Rechenanlage des Astronomischen Rechen-Instituts.

E. Høg: The photoelectric micrometer for the Bergedorf meridian circle.

I. Torgård: Stellar orbits in a three dimensional galactic model system.

H.G. Groth: Electronic computation of stellar model atmospheres.

A. Elvius: Photoelectric observations of polarization in M 82 and other galaxies.

W. Seitter: Zwei-Farben-Diagramme in verschiedenen photometrischen Systemen.

K. Gyldenkerne: On the narrow band classification of G and K giants.

T. Ringnes: Long period variations in sun spot activity.

**8  Meeting in Hamburg 1967**

The meeting was called "Baltic Neighbourhood Conference Hamburg" and took place in Hamburg 19.-20. May 1967, with informal gathering at Restaurant Curiohaus on Thursday, May 18. The sessions were held in "Hörsaal B – Philosophenturm" in the University Campus.

The program signed by G. Traving is provided at Hamburg (1967) and lists the speakers and titles. In sequence the speakers were: K. Gyldenkerne (invited) – Brorfelde, T. Oja – Uppsala, K. Johansen – Kopenhagen, B. Helt – Kopenhagen, C. Roslund – Lund, P.E. Nissen – Arhus, B. Bashek (invited) – Kiel, K. Tonner – Kiel, A. Behr – Göttingen, A. Weigert – Göttingen, J. Baerentzen – Arhus, K. Fricke – Göttingen, W. Dieckvoss (invited) – Hamburg, N. Hansson – Lund, Fr. von Fischer-Treuenfeld – Hamburg, S. Laustsen – Kopenhagen, H. Haffner – Hamburg/Würzburg.

Three invited contributions of 45 min were placed at the beginning of each half-day session. The subjects were, as recorded in the Hamburg Jahresbericht, "empirical and theoretical problems of narrow-band photometry and astrometric issues":

"Narrow-band photometry and its empirical aspects" by K. Gyldenkerne

"Theoretical interpretation of narrow-band photometry" by B. Baschek

"New contributions of the Hamburg Observatory to proper motion work in the northern hemisphere in the international program AGK3" by W. Dieckvoss.

For the first evening "our colleagues are kindly invited to refresh their acquaintance with German wines – at about 20 p.m. – in the library of the Observatory in Bergedorf."

The program ends such: "The Hamburg Observatory will be open for all guests on Saturday Afternoon. Drs. D. Rudolph and G. Schmahl (Göttingen) will take the opportunity to demonstrate gratings and zone plates made by a new method."

**9  Correspondence**

Information on the Baltic Meetings of astronomers from Northern Germany and Scandinavia from Bonn and Göttingen to Uppsala has been received from Bengt Edvardsson (Uppsala), Svend Laustsen (Aarhus, retired), Lennart Lindegren (Lund), Peter Naur (Copenhagen, retired), and Jørgen Otzen Petersen (Copenhagen, retired) and something was found in my own memories and archive. This information includes the programs of the meetings and covers the subject quite well. Information has been requested by mail in January and February 2015 to about twenty persons related to the institutes in Bonn, Göttingen, Hamburg, Kiel, Copenhagen, Aarhus, Lund, Oslo, Stockholm and Uppsala.

Up to the deadline set at 2015-04-10, responses were received from the following: Michael Geffert in Bonn informs that they "don't have a real archive"; Wilhelm Kegel formerly in Kiel thought that Charlotte Schönbeck could have something, but she denies; Anke Vollersen in Hamburg informs that no documents or photos from the meetings have been found, only brief notes in the Jahresbericht. Kaare Aksnes in Oslo could not add new information. Wolfram Kollatschny in Göttingen sent a page from the guest book. Axel Wittmann in Göttingen informs that from Otto Heckmann only reprints on cosmology have come down, nothing about Baltic meetings. Responses have been received from Sebastian Wolf in Kiel, and Per Olof Lindblad in Stockholm, but neither had new information.

**10  Conclusions**

The Baltic meetings around 1960 played a role in times when such rather local gatherings were seldom and travelling was fairly complicated and expensive. Specialized and more international meetings soon took over.

We may consider the astronomical interests 50 years ago based on the number of presentations in the meetings in Lund, Göttingen and Copenhagen. The ratio of astrometry to astrophysics was 4 to 11 in Lund, 1 to 19 in Göttingen, and 3 to 6 in Copenhagen. The high ratios for astrometry at the meetings in Lund and Copenhagen would be unthinkable today in a general astronomical audience, but that was how my generation of astronomers grew up in Europe, the generation which was strong enough to get space astrometry started and carried through. I do not at all regret the change of times, my only concern is that the astrometric community could suffer a decline after the Gaia mission such that the capability to support a new large astrometry mission in Europe could be lost. Therefore I am arguing that a new global astrometry mission by the European Space Agency in 20 years is required to secure an astrometric foundation of astrophysics of the highest quality (Høg 2014b, c).

Another conclusion from my correspondence concerns archives from these years, the decade around 1960, at the ten observatories in Copenhagen, Aarhus, Lund, Stockholm, Uppsala, Oslo, Kiel, Hamburg, Göttingen and Bonn. Practically nothing exists from any of the last nine places, according to the correspondence documented in Høg (2015). An explanation may be that give by one of my correspondents who wants to remain anonymous: "The preservation in archives is always largely a matter and responsibility of the directors of the institutes, who normally do not have an education in history, law etc. as they should have. ... I have also discarded most of my ... documents because their original purpose has been fulfilled."

From the Copenhagen Observatory about 7000 letters and other correspondence of administrative and scientific character dated between 1947 and 1959 are presently placed at the Kroppedal Museum, near Copenhagen. This is briefly listed in section 3 of Høg (2015). A selection of other letters, mostly in Danish, from Bengt Strömgren and Julie Vinter Hansen, and a few from Peter Naur, Otto Heckmann, and Erik Høg are listed in section 2 of Høg (2015) where links to some of the letter are given. Finn Aaserud, director of the Niels Bohr Archive, Copenhagen, has emphasized the "great importance of keeping this documentation permanently available for historians [in paper form]."

**Acknowledgements:** I am grateful to Gudrun Wolfschmidt for the invitation to contribute to the meeting "Astronomie im Ostseeraum" and to Finn Aaserud, Kaare Aksnes, Bengt Edvardsson, Michael Geffert, Bodil Helt, Wilhelm Kegel, Wolfram Kollatschny, Svend Laustsen, Per Olof Lindblad, Lennart Lindegren, Peter Naur, Lars Occionero, Jørgen Otzen Petersen, Charlotte Schönbeck, Anke Vollersen, Axel Wittmann, and Sebastian Wolf for discussion and information.

## 11  References

A note on links for the reader of a printed version: The present report is placed at
http://www.astro.ku.dk/~erik/xx/BalticMeetings.pdf where the internet links to most references are therefore directly available.

Copenhagen 1962, List of participants in Copenhagen in April 1962.
http://www.astro.ku.dk/~erik/xx/Copenhagen1962.pdf

Hamburg 1967, Baltic Neighbourhood Conference Hamburg. Program signed by G. Traving.
http://www.astro.ku.dk/~erik/xx/BalticNeighbourhoodConferenceHamburg1967.pdf

Høg E. 2008, Lennart Lindegren's first years with Hipparcos. www.astro.ku.dk/~erik/Lindegren.pdf Also available as report no. 2 in "Astrometry and optics during the past 2000 years". November 2008, 8+94 pages, a collection of reports at http://arxiv.org/abs/1104.4554

Høg E. 2011a, Astrometry Lost and Regained - From a modest experiment in Copenhagen in 1925 to the Hipparcos and Gaia space missions. http://www.astro.ku.dk/~erik/AstromRega3.pdf. Also available as report no. 10 in "Astrometry during the past 100 years". April 2011, 8+46 pages, a collection of reports at http://arxiv.org/abs/1105.0634

Høg E. 2011b, Astrometry history: Roemer and Gaia. http://arxiv.org/abs/1105.0879

Høg E. 2014a, Astrometry 1960-80: from Hamburg to Hipparcos. Proceedings of conference held in Hamburg in 2012, Nuncius Hamburgensis, Beiträge zur Geschichte der Naturwissenschaften, Band 24, 2014. http://arxiv.org/abs/1408.2407

Høg E. 2014b, The Astrometric Foundation of Astrophysics. Abstract to the Conference Book 2014 of the Danish  Astronomical Society and abstract of a review presentation.  http://arxiv.org/abs/1408.2122

Høg E. 2014c, Absolute astrometry in the next 50 years. http://arxiv.org/abs/1408.2190

Høg E. 2015, Archives on astronomy from the 1950s.
 http://www.astro.ku.dk/~erik/xx/Archive1950s.pdf

NAT 1962, Astronommøde I København. In: Danish in Nordisk Astronomisk Tidsskrift (1962), p. 100, 4 pages in Danish: http://www.astro.ku.dk/~erik/xx/astronomy_meeting_Copenhagen_1962.pdf

Naur, P. 2015, Note on controversies on astronomical aberration.
http://www.astro.ku.dk/~erik/xx/NaurAberration.pdf

PAT 1956, Astronomiskt symposium I Lund. In: Populär Astronomisk Tidskrift 38, 155 (1957), 1 page in Swedish.  http://www.astro.ku.dk/~erik/xx/astronomy_meeting_Lund_1957.pdf